\documentclass[aps,pra,twocolumn, 10pt, showpacs,superscriptaddress,groupedaddress]{revtex4-1}  

\newcommand{\ket}[1]{\left\vert#1\right\rangle}

\newcommand{\braket}[1]{\left\langle#1\right\rangle}

\newcommand{\Rb}{\ensuremath{^{87}\text{Rb }}}

\newcommand{\eq}[2][{}]{\begin{equation}\label{#1} #2 \end{equation}}

\newcommand{\ie}{\emph{i.e.} }

\newcommand{\N}{\ensuremath{\mathcal{N}}}

\usepackage{graphicx}
\usepackage{epstopdf}
\usepackage{setspace}
\usepackage{bbold}
\usepackage{comment}
\usepackage{color}
\usepackage{soul}
\usepackage[abs]{overpic}
\usepackage{amsmath}
\usepackage{lipsum}
\usepackage[abs]{overpic}
\usepackage{amsmath}
\begin{document}

\author{Gadi Afek}
\affiliation{Department of Physics of Complex Systems, Weizmann Institute of Science, Rehovot 76100, Israel }
\author{Jonathan Coslovsky}
\affiliation{Department of Physics of Complex Systems, Weizmann Institute of Science, Rehovot 76100, Israel }
\author{Alexander Mil}
\altaffiliation[Current address: ]{Kirchhoff-Institut f\"ur Physik, Ruprecht-Karls-Universit\"at Heidelberg, Im Neuenheimer Feld 227, 69120 Heidelberg, Germany }
\author{Nir Davidson}
\affiliation{Department of Physics of Complex Systems, Weizmann Institute of Science, Rehovot 76100, Israel }

\title[]{Revival of Raman Coherence of Trapped Atoms}

\begin{abstract}
We perform Raman spectroscopy of optically trapped non interacting \Rb atoms, and observe revivals of the atomic coherence at integer multiples of the trap period. The effect of coherence control methods such as echo and dynamical decoupling is investigated experimentally, analytically and numerically, along with the effect of the anharmonicity of the trapping potential. The latter is shown to be responsible for incompleteness of the revivals. Coherent Raman control of trapped atoms can be useful in the context of free-oscillation atom intefrerometry and spatial multi-mode quantum memory.
\end{abstract}

\maketitle

Coherent control of cold neutral atomic ensembles can be used for a variety of studies and applications in physics, such as metrology, quantum information, quantum simulators, and fundamental quantum mechanics~\cite{Loh2013,Zhang2016,Asenbaum2017,Cairncross2017,Riofrio2017,Mazurenko2017}. Long coherence times usually imply the need to minimize perturbations on the atoms, as implemented in atomic fountain clocks and atom interferometers. However, in some cases, such as quantum memory~\cite{Duan2001,Lukin2003,Gorshkov2007,Moretti2008,Zhao2009Long,zhao2009millisecond,Heinze2013}, there arises also a need for long coherence times in trapped samples. This approach involves challenges such as the inhomogeneous profile of both the external potential and the atomic density~\footnote{in part alleviated by use of homogeneous trapping potentials~\cite{Gaunt2013,Mukherjee2017}} and elastic atomic collisions. Coherence has also been shown to revive, in the deep quantum regime, via interaction mechanisms in a Bose-Einstein condensate~\cite{Greiner2002,Egorov2011}.

Quantum control of the internal state of a trapped atom can be achieved by the use of microwave (MW) radiation, allowing extremely long coherence times~\cite{Cornell2002,Treutlein2004,Sagi2010_Universal,Deutsch2010,Kleine2011,Coslovsky2017Collisions} and enabling the implementation of atom-chip clocks and single-pixel quantum memories. 

There are, however, applications for which MW control does not suffice. In two-photon stimulated Raman control the momentum recoil is given by $\hbar k_{\text{eff}} \approx 2 \hbar k \sin{(\alpha/2)}$, where $\hbar$ is the reduced Planck constant, $k$ is the wave number of the Raman beams and $\alpha$ is the angle between them. By changing the angle $\alpha$, the momentum recoil $\hbar k_{\text{eff}}$ can be varied between practically zero and $2\hbar k$. This allows coupling to the external degrees of freedom of the atoms and the possibility of implementation of spatial multi-mode quantum memory~\cite{Shuker2008,Nicolas2014}.

Raman atomic coherence in a trap is closely related to the fringe contrast of guided interferometers~\cite{Dumke2002,Burke2008,McDonald2013,McDonald2013_2}, and more specifically free-oscillation atom interferometers~\cite{Horikoshi2007,Burke2008,Sapiro2009,Segal2010,Kafle2011,Leonard2012,Fogarty2013}. These rely on the classical turning points of an underlying harmonic potential for the mirroring of the wave packets. A thermal atom free-oscillation Raman interferometer allows for Ramsey $\pi/2\to\pi/2$ interferometry which is completely impossible to perform in free-space, utilizing the periodicity of the trapping potential. This type of interferometer is sensitive to time-dependent forces~\cite{Kafle2011,Coslovsky2017}, whereas the addition of a $\pi$ (echo) pulse in between the Ramsey pulses allows for sensitivity to DC accelerations.

In this paper we demonstrate experimentally, analyze theoretically and validate numerically the revivals of the Raman coherence due to the effect of the trapping potential. We show that for a dipole-trapped ensemble of non interacting \Rb atoms, the coherence decays at a rate which depends on the momentum transferred by the Raman beams and the temperature of the ensemble. It then revives at integer multiples of the trapping period. Anharmonicity of the trapping potential is found to deteriorate the amplitude of the revival. We further show that by using quantum control methods, such as the Hahn-echo and dynamical-decoupling techniques with the Raman beams, the deterioration due to anharmonicity can be reduced and the coherence kept high. 

\subsection*{Raman coherence in a harmonic potential}

The coherence of an atomic ensemble can be accessed via a Ramsey experiment. Two $\pi/2$ pulses are applied to atoms initially in the ground state $\ket{1}$ of a two-level system, with a time delay $t$ in between. Immediately after the first pulse the internal atomic state in the rotating frame is given by $\ket{\psi} = \frac{1}{\sqrt{2}}\left(\ket{1}+ e^{i \vec{k}_{\text{eff}} \cdot \vec{r}_0} \ket{2}\right)$, where $\vec{k}_{\text{eff}} = \vec{k}_1-\vec{k}_2$, and $\vec{r}_0$ is the initial position of the atom. The Ramsey coherence upon readout is given by
\eq{C(t) = \left| \left< e^{i \vec{k}_{\text{eff}} \cdot \left[\vec{r}_0 - \vec{r}(t)\right]+ i \phi(t)} \right> \right|, 
\label{eq:coherence_no_pi}}
where $\vec{r}(t)$ is the atomic position at time $t$, and $\phi$ is the phase accumulated due to the differential light shift of the trapping laser~\footnote{\Rb atoms trapped in an optical dipole trap experience a differential AC-Stark shift imposed by the different detuning of the trapping laser from their two ground state hyperfine levels. Effectively this creates a stationary inhomogeneous broadening of the spectrum, decreasing the coherence time of the ensemble~\cite{Kuhr2005}.}.

When the dephasing is dominated by the Doppler shift of the Raman beams, Eq.~\eqref{eq:coherence_no_pi} can be separated to a product of the MW and Raman coherences
\eq{\begin{split} C(t) &= \left| \left< e^{i \phi(t)} \right> \right| \times \left| \left< e^{i \vec{k}_{\text{eff}} \cdot \left[\vec{r}_0 - \vec{r}(t)\right]} \right> \right| \\
& = C_\text{MW}(t) \times C_{\text{R}}(t),\end{split}\label{eq:split_coherence}}
where $C_\text{R}(t)$ is the fast evolving coherence due to the Raman control, and $C_\text{MW}(t)$ is the slow evolving coherence due to all other factors. The properties of $C_\text{MW}(t)$ have been extensively studied~\cite{Cornell2002,Treutlein2004,Sagi2010_Universal,Deutsch2010,Kleine2011,Coslovsky2017Collisions}. Typically, and in particular under our experimental conditions, $C_\text{MW}$ decays much slower than $C_\text{R}$, rendering its dynamics negligible.

For non-interacting atoms trapped in a 1D harmonic trap of frequency $\omega$~\footnote{For non-interacting particles in a perfectly harmonic trap the generalization to 3D is trivial, as the different axes are uncoupled.}, the trajectory of each atom is given by $x(t) = x_0 \cos(\omega t) + \frac{v_0}{\omega}\sin(\omega t)$, where $x_0,v_0$ are its initial position and velocity. The Ramsey coherence can be analytically derived by substituting this into $C_\text{R}$ of~Eq.~\eqref{eq:split_coherence} and averaging over the Boltzmann-distributed initial conditions $f(v_0,x_0)\sim \exp\left[-m(v_0^2+\omega^2 x_0^2)/2k_BT\right]$. Here $m$ is the atomic mass, $k_B$ is the Boltzmann constant and $T$ is the temperature of the atomic ensemble. The resultant coherence is given by
\begin{equation}
C_\text{R}(t) = \exp\left[-\N^2 \left[1-\cos(\omega t)\right]\right],
\label{eq:coherence_harmonic}
\end{equation}
where we have defined the number of phase fringes on the atomic cloud as $\N = k_{\text{eff}} x_{\text{RMS}}$, with $x_{\text{RMS}} = \sqrt{\frac{k_B T}{m \omega^2}}$ the size of the atomic cloud. Figure~\ref{fig:fig1}(a) shows a plot of Eq.~\eqref{eq:coherence_harmonic} for three values of $\N$, revealing that whenever $\omega t$ is an even multiple of $\pi$ there is a full revival of coherence. Minimal coherence, $e^{-2\N^2}$, is reached whenever $\omega t$ is an odd multiple of $\pi$. For $\omega t \ll 1$ the coherence decays as a Gaussian $\sim \exp\left[-(t/\tau)^2\right]$, with $\tau = \sqrt{2}/\N\omega = \sqrt{2}/k_{\text{eff}} v_\text{RMS}$, where $v_\text{RMS} = \sqrt{k_BT/m}$ is the RMS velocity. In the limit $\omega t \rightarrow 0$, the initial decay rate converges to that of free atoms up to a factor of $\sqrt{2}$~\footnote{The reason for the $\sqrt{2}$ discrepancy is that under thermal equilibrium, the equipartition theorem dictates that as $\omega t \rightarrow 0$, the size of the atomic cloud grows and even at the limit there are always atoms experiencing strong potential gradients and the effect of the harmonic potential cannot be neglected.}.

\subsection*{Effects of anharmonicity}

\begin{figure}
\centering
\begin{overpic}
[width=\linewidth]{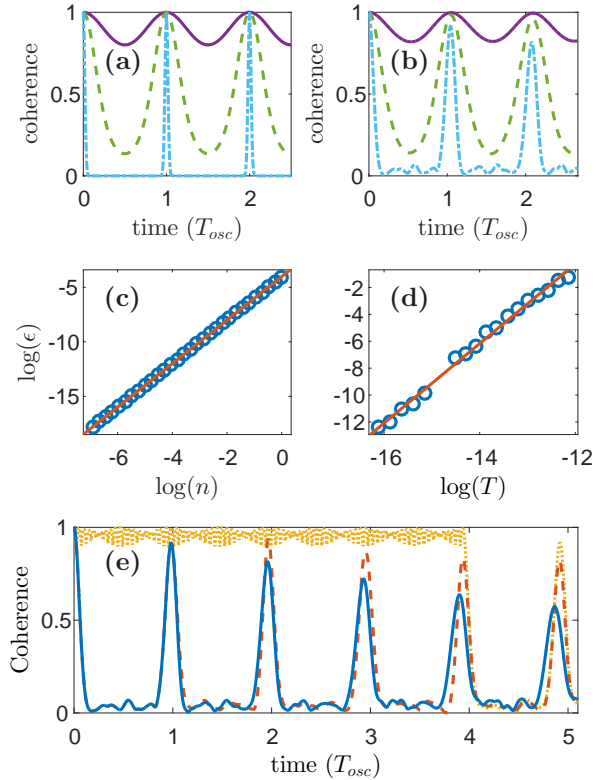}
\put(40,280){ \textbf{(a)}}
\put(148,280){ \textbf{(b)}}
\put(40,190){ \textbf{(c)}}
\put(148,190){ \textbf{(d)}}
\put(40,90){ \textbf{(e)}}
\end{overpic}
\caption[Ramsey coherence in a trap.]{{Simulated Ramsey coherence in a trap.} {\bf (a)} Plot of the result of Eq.~\eqref{eq:coherence_harmonic} for $\N=1/3$ (purple, solid), $\N=1$ (green, dashed) and $\N=3$ (light blue, dash-dotted). The coherence drops rapidly at an $\N$-dependent timescale and revives fully after an oscillation period in the trap. {\bf (b)} In contrast to the perfectly harmonic trap case of (a), for a Gaussian trap (with anharmonicity $\mathcal{A}=0.025$) the simulated revivals do not reach unity coherence, but rather decrease monotonically. {\bf (c)} Scaling of $\epsilon$, the infidelity of the revival [Eq~\eqref{eq:error}], with $\N$ gives a slope of 1.993~(2), in agreement with prediction. This simulation is a single run at a given temperature in which we scan $\N$ by changing the angle between the Raman beams. {\bf (d)} $\log (\epsilon)$ as a function of $\log (T)$. The linear fit gives a slope of 2.96~(4), also in agreement with the prediction. {\bf (e)} An improvement over the anharmonic-trap Raman coherence (solid blue line) is achieved by applying echo pulses at the peaks of the revivals (dashed red line), or by using dynamical decoupling (dotted orange line). In the dynamical decoupling simulation the rate of $\pi$-pulses is 40 pulses per $T_{\text{osc}}$. The dynamical decoupling pulses are applied until $t=4T_{\text{osc}}$, at which point the coherence drops and revives after yet another oscillation period. All errors correspond to a $1\sigma$ confidence level.}
\label{fig:fig1}
\end{figure}

For a thermal cloud in a symmetric trap that is not perfectly harmonic, each atom has a slightly different oscillation period and therefore at time $t = T_\text{osc}= 2 \pi/\braket{\omega}$, the atoms do not return to their initial position, reducing the revival amplitude. We perform a Monte Carlo simulation, solving the equations of motion for $10^3$ atoms in a 3D Gaussian confining potential and calculating their coherence according to Eq.~\eqref{eq:coherence_no_pi}. Fig.~\ref{fig:fig1}(a-b) compare the analytic result of Eq.~\eqref{eq:coherence_harmonic} to simulation results for different values of $\N$, revealing good agreement between the two. In the simulations [Fig.~\ref{fig:fig1}(b)], however, the effect of anharmonicity is manifested in a broadening and a reduction in amplitude of the Raman revivals as $\N$ increases.

To further analyze this effect we define a dimensionless ensemble anharmonicity $\mathcal{A}\equiv\braket{\omega}/\omega-1$, under the assumption that the oscillation frequency shift of each atom is proportional to its initial energy. Here $\omega$ is the (harmonic) frequency at the trap bottom. For a perfectly harmonic trap $\mathcal{A} = 0$, and for our crossed Gaussian-beam dipole trap $\mathcal{A} < 0$. Calculating the phase at integer multiples $\nu$ of the trapping period $T_\text{osc}$ to leading order in $\mathcal{A}$ yields the error in the $\nu^\text{th}$ revival caused by anharmonicity 
\begin{equation}\label{eq:error}
    \epsilon_\nu\equiv1-C_\text{R}\left(t=\nu T_\text{osc}\right)\approx 6\pi^2\mathcal{A}^2\N^2\nu^2.
\end{equation}
We test this scaling by comparing it to simulations with a constant anharmonicity $\mathcal{A}=0.025$ and varying $\N$ [Fig.~\ref{fig:fig1}(c)]. The fitted slope in a log-log scale is 1.993~(2), very close to 2, as expected. This scaling prediction can also be expanded to other experimental parameters, such as temperature. Since $\left\vert\mathcal{A}\right\vert \sim T$ and $\N \sim T^{1/2}$, we get that $\epsilon \sim T^3$~\footnote{The scaling for $\mathcal{A}$ comes from the assumption that the oscillation frequency shift of each atom is proportional to its initial energy. The scaling for \N is a consequence of the equipartition theorem.}. We test this prediction by comparing simulation results at different temperatures to the expected scaling law [Fig.~\ref{fig:fig1}(d)]. The fitted slope in the log-log scale is 2.96~(4), in agreement with the prediction. 

A well established method for extending the coherence is the Hahn echo ($\pi/2\to\pi\to\pi/2$)~\cite{Hahn}, where the inverting $\pi$ pulse is applied at time $t_\pi$. Neglecting global phases, the effect of the $\pi$-pulse is given by $\ket{1} \rightarrow \ket{2}$ and $\ket{2} \rightarrow e^{-2i \vec{k}_{\text{eff}} \cdot \vec{r}(t_\pi)}\ket{1}$. The coherence is then
\eq{C_\text{R}(t>t_\pi) = \left| \left< \exp\left[i \vec{k}_{\text{eff}} \cdot \left[2\vec{r}(t_\pi)-\vec{r}_0 - \vec{r}(t)\right]\right] \right> \right|
\label{eq:coherence_Echo}.}
In a perfectly harmonic trap, when $t_\pi = 2\pi/\omega = T_\text{osc}$, the oscillation period in the trap, the echo-pulse has no effect on the coherence. According to our simulations, when echo-pulses are applied at the times of coherence revival in an anharmonic trap, the following revivals improve. This is because to first order the phase vanishes independently of the displacement of the atoms (which differs from atom to atom), and the echo pulse overcomes the anharmonicity [Fig.~\ref{fig:fig1}(e)]. Taking into account second order effects, we find $\epsilon \sim \mathcal{A}^4 \N^2$. On the other hand, if $t_\pi = T_\text{osc}/2$, 
\eq{C_\text{R}(t>t_\pi) = \exp\left[-\N^2\left[5+3\cos{(\omega t)}\right]\right]. \label{eq:bad_echo}}
The coherence drops to $e^{-8\N^2}$ at $t=T_\text{osc}$, and experiences a small revival at $t=3T_\text{osc}/2$ to a value of $e^{-2\N^2}$. An echo pulse applied at $t=T_\text{osc}$ keeps the coherence high at the time of the following revival because the dynamics before and after the pulse are nearly identical. An echo-pulse applied at $t = T_\text{osc}/2$ dramatically decreases the following coherence due to the fact that here dynamics before and after the pulse are opposite.

A more robust method for reducing anharmonicity-driven decoherence is dynamical decoupling (DD)~\cite{Biercuk2009,Almog2011,Genov2017}. Application of a fast sequence of $\pi$-pulses (\ie many pulses within the trap oscillation period) can be described by a generalization of Eq.~\eqref{eq:coherence_Echo}, yielding the result depicted in orange in Fig.~\ref{fig:fig1}(e). The coherence remains high up to $t = 4 T_{\text{osc}}$, which is when the simulated DD pulses stop. Then there is a decay of the coherence followed by a revival at $t = 5 T_{\text{osc}}$. 

DD can also be useful in mitigating the effect of interactions~\cite{Sagi2010_Process}. A prominent example of such interactions is elastic atomic collisions occurring at a rate $\Gamma_\text{coll}$, which can hinder the coherence significantly. A single collision will reduce both the Ramsey and echo revivals because when an atom collides, its trajectory is changed and therefore it does not return to its original position at $t=T_{\text{osc}}$. Under the effect of a set of DD pulses given at a frequency $f_\text{DD}$, the attainable coherence time, $t_c$, can be derived by adapting the calculation of~\cite{Sagi2010_Process}. It is the time for which the width of phase distribution is on the order of the effective wavelength of the Raman control $\lambda_\text{eff}\sim 1/k_\text{eff}$, yielding $t_c\sim \tau_D^2f_\text{DD}^2/\Gamma_\text{coll}$ with the notable quadratic dependence on $f_\text{DD}$. The Doppler time $\tau_D=v_\text{RMS}k_\text{eff}$ is the RMS velocity of the cloud times the effective Raman wave number.

\subsection*{Experimental setup and results}

Our cold atoms apparatus is described in detail in \cite{Sagi2010_Universal}. In short, it consists of $6\times 10^4$ $^{87}$Rb atoms at a temperature of about $2.4 \mu$K trapped in a crossed optical trap. Trapping frequencies are $(\omega_x, \omega_y, \omega_z) \approx 2 \pi \times (385, 385, 110)$~Hz giving $x_\text{RMS} \approx 6~\mu$m. The Raman transitions are performed between the $\ket{1} \equiv \ket{F=1,m_F=0}$ and $\ket{2} \equiv \ket{F=2,m_F=0}$ first-order Zeeman insensitive states of the $5^2S_{1/2}$ manifold. At the beginning of each experiment the atoms are prepared in the $\ket{1}$ state. At the end of each experiment we use a state-selective fluorescence-detection (``normalized detection'') scheme to evaluate the fraction of atoms at state $\ket{2}$. 

For the Raman control we use a distributed feedback laser offset-locked to an atomic reference~\cite{offset_locking_paper}, yielding a one-photon detuning of -3.4~GHz relative to the $F'=2$ level of the $D_1 (5^2P_{1/2})$ transition at a wavelength of 795~nm. The output beam is split into two, one directly transported to the atoms via single-mode, polarization maintaining optical fiber and the other passing a fiber-coupled electro-optic modulator before being transported to the atoms in the same way. A 10~MHz frequency shift is used to select only one of the sidebands created by the electro-optic modulator. 

Each beam hits the trapped atomic cloud at a small angle ($\pm \alpha/2$) from the trap longitudinal symmetry axis. A magnetic field corresponding to 100~kHz in the direction of this symmetry axis sets the quantization axis. The two beams are circularly polarized to give a large $\sigma^+$ component, only allowing transitions that preserve $m_F$. By setting the two-photon detuning, only the $\ket{1} \leftrightarrow \ket{2}$ transition is resonant.

\begin{figure}
\centering 
	\begin{overpic}
[width=\linewidth]{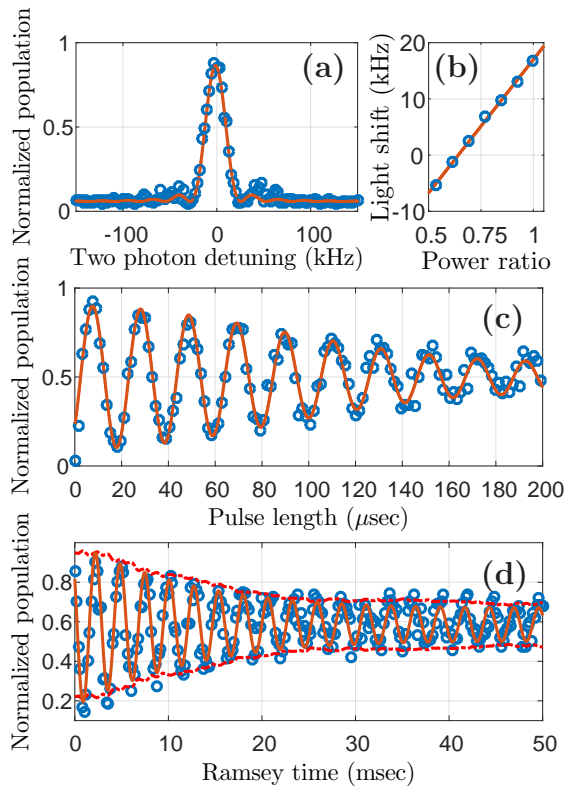}
\put(130,275){\large \textbf{(a)}}
\put(180,275){\large \textbf{(b)}}
\put(198,180){\large \textbf{(c)}}
\put(198,83){\large \textbf{(d)}}
\end{overpic}
\caption[Raman Ramsey experiment]{{Doppler-free Raman spectroscopy ($\alpha\approx 0.1^{\circ}$).} {\bf (a)} Atomic Rabi spectrum. The atoms perform the $\ket{F=1,m_F = 0}\to\ket{F=2,m_F = 0}$ transition indicated by the measurement of the normalized $\ket{F=2,m_F = 0}$ population (vertical axis). Horizontal axis is the detuning from the two-photon resonance of $\approx6.8~$GHz. {\bf (b)} Light shift induced by the Raman beams as a function of the ratio between their powers. By properly choosing the detuning and the power ratio, the light shift can be either canceled out or increased. Solid line is a linear fit. {\bf (c)} Rabi Oscillations. The two-photon Rabi frequency is 49~kHz (solid line is a fit to an exponentially decaying sine) and the decay time is 130 $\mu$s, dominated by inhomogeneity in the Raman beams. The data presented is an average over three realizations. {\bf (d)} Ramsey fringes. The extracted envelope (dash-dotted red line) shows long coherence times. Solid line is a fit to an exponentially decaying sine. Normalized detection stands for the fraction of atoms in the $\ket{F=2}$ state.}
\label{fig:fig2}
\end{figure}

In Fig.~\ref{fig:fig2}, we present a set of$\alpha \approx 0.1^\circ$, practically Doppler-free spectroscopic measurements obtained using our Raman control. Fig.~\ref{fig:fig2}(a) presents a Rabi spectrum. Fitting the obtained spectrum to a squared sinc function we find that the resonance frequency is shifted from that of the hyperfine transition by about 1.4~kHz. This is mainly due to the differential light shift caused by the Raman control itself. For our chosen one-photon detuning, the Raman beams generate light shifts of opposite signs, meaning that by properly choosing the power ratio between them, the shift can be nullified [Fig.~\ref{fig:fig2}(b)]. This ability to control the light shift enables us to have a controlled detuning without having to sacrifice the quality of the pulses and the contrast of the Rabi fringes, providing rapid Ramsey oscillations and allowing exploration of the fast decay dynamics. Fig.~\ref{fig:fig2}(c) shows Raman Rabi oscillations, decaying to $1/e$ within about 6.4 oscillations (130~$\mu$s), dominated by spatial inhomogeneity of the Raman control which causes different atoms to experience a different effective Rabi frequency. This is sufficient for few-pulse sequences such as Ramsey and echo, but a limiting factor in the ability to perform many-pulse sequences such as DD. Fig.~\ref{fig:fig2}(d) presents a negligible-angle Raman Ramsey measurement, imitating the dynamics of the MW coherence. Represented by the contrast of the fringes, the coherence decays on a time scale on the order of tens of milliseconds, comparable to decay times obtained with similar experimental conditions using MW control~\cite{Sagi2010_Universal}.  

\begin{figure}
\centering 
	\begin{overpic}
[width=0.9\linewidth]{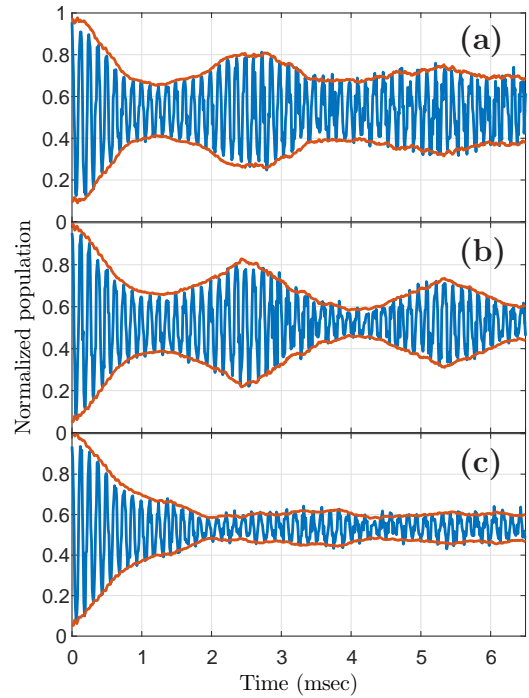}
\put(175,250){\large \textbf{(a)}}
\put(175,170){\large \textbf{(b)}}
\put(175,90){\large \textbf{(c)}}
\end{overpic}\\
\caption[Ramsey and echo with Raman beams]{{Raman Ramsey and echo.} {\bf (a)} For a finite, small angle ($\alpha \approx 0.9^{\circ}$) a rapid partial decay is observed, followed by clear partial revivals at $t = T_{\text{osc}}$ and $t = 2T_{\text{osc}}$. {\bf (b)} An echo experiment with a $\pi$-pulse given at $t = T_{\text{osc}}$. The contrast of the revivals is augmented. {\bf (c)} An echo experiment with a $\pi$-pulse given at $t = T_{\text{osc}}/2$. As a result the revival at $t = T_{\text{osc}}$ disappears and a small revival appears at about $t = 1.5T_{\text{osc}}$. Red lines, summarized in Fig.~\ref{fig:fig4}(a), depict the coherence, extracted by taking the standard deviation over an oscillation.}
\label{fig:fig3}
\end{figure}

Fig.~\ref{fig:fig3}(a) shows a Ramsey measurement induced by the Raman beams separated by an angle $\alpha \approx 0.9^\circ$, generating $\N \approx 0.78$ fringes on the cloud. The coherence decays at a timescale $\tau\approx 780~\mu$s, compared to the expected value of $\tau = \sqrt{2}/\N\omega \approx 750~\mu$s calculated from the independently measured temperature and trap oscillation frequencies. This is much faster than that of the small-angle Fig.~\ref{fig:fig2}(d), and justifies the neglection of $C_\text{MW}$ in the derivation of Eq.~\eqref{eq:split_coherence}. The coherence decays to a minimum value of 0.29, in agreement with the predicted value of 0.3 obtained from Eq.~\ref{eq:coherence_harmonic}. A clear revival can be seen around $t = T_{\text{osc}} = 2.6$~ms. The revival amplitude is 0.61. Fig.~\ref{fig:fig3}(b) presents an echo experiment in which the echo $\pi$-pulse is given at $t_\pi = T_{\text{osc}}$. The second revival, at $2T_{\text{osc}}$, becomes sharper. The echo pulse, however, does not improve the coherence of the sequential revival. Fig.~\ref{fig:fig3}(c) shows an echo experiment with a $\pi$-pulse given at $t \approx T_{\text{osc}}/2$, predicted to cause a deterioration of the coherence. As expected from the analysis of Eq.~\eqref{eq:bad_echo}, a rapid decrease of the coherence is observed after the echo-pulse with a small revival at $t = 1.5T_{\text{osc}}$. 

\begin{figure}
\centering 
\begin{overpic}
[width=\linewidth]{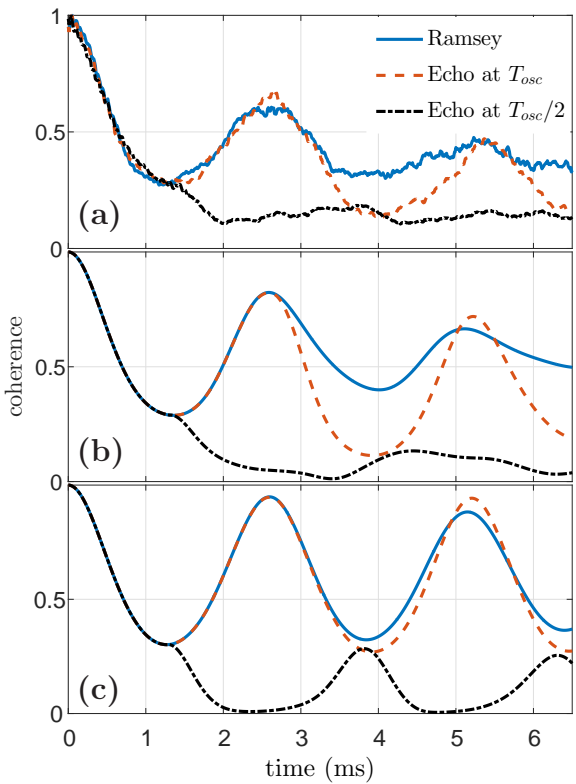}
\put(35,215){\large \textbf{(a)}}
\put(35,125){\large \textbf{(b)}}
\put(35,35){\large \textbf{(c)}}
\end{overpic}
\caption[Echo]{{Echo dynamics.} {\bf (a)} A summary of measurements (a-c) of Fig.~\ref{fig:fig3}. {\bf (b)} Simulated coherence for a trap with measured anharmonicity of $\mathcal{A}=0.4$ (see appendix), in excellent agreement with the experiment with no fitting parameters. {\bf (c)} Simulation results for an ideal crossed Gaussian beam trap, highlighting the deteriorating effect of the added anharmonicity on the atomic coherence.}
\label{fig:fig4}
\end{figure}

Fig.~\ref{fig:fig4}(a) summarizes the results of the Ramsey and echo experiments, where the echo pulse is given at both $T_\text{osc}$ and $T_\text{osc}/2$ [Fig.~\ref{fig:fig3}(a-c)], normalized to the natural scale of $C_\text{R}\in[0,1]$, and compares them to Monte Carlo simulations for a realistic trap (see appendix) and for an ideal Gaussian cross-beam trap. Fig.~\ref{fig:fig4}(b) presents the simulated coherence for the trap with the measured anharmonicity, in excellent agreement with all of the experimental data and without any free fitting parameters. Fig.~\ref{fig:fig4}(c) shows results for the ideal crossed beams Gaussian trap, highlighting the effect of the added anharmonicity. 

\begin{figure}
	\centering
\begin{overpic}
[width=\linewidth]{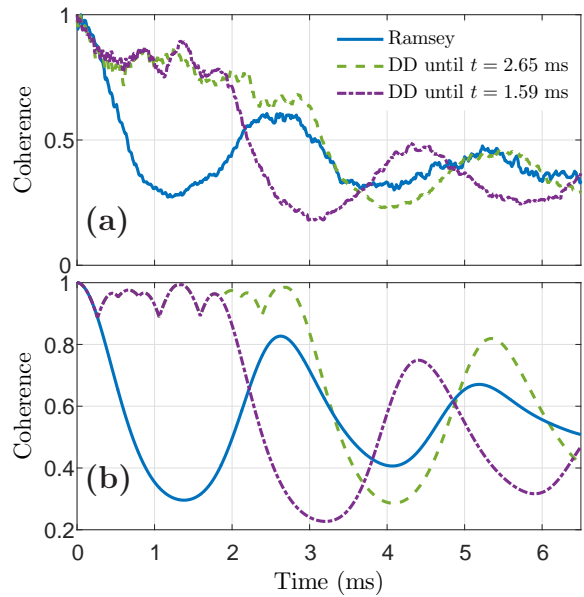}
\put(35,140){\large \textbf{(a)}}
\put(35,42){\large \textbf{(b)}}
\end{overpic}
		\caption[Dynamical decoupling]{{Dynamical decoupling.} {\bf (a)} Experimental results. During the DD, the coherence remains high. A revival appears about $T_{\text{osc}}$ after the DD has ended. The number of $\pi$-pulses is 10 and 6 for the 2.65~ms and 1.59~ms data respectively. {\bf (b)} Simulation results in a calibrated anharmonicity trap.}
\label{fig:fig5}
\end{figure}

Fig.~\ref{fig:fig5} shows results of a DD experiment (a) and compares them to simulation (b). As long as the DD is active the coherence remains high, up to a slow decay appearing in the experimental data and not in the simulation. This is due to the finite number of Rabi oscillations available in the experiment before loss of coherence [Fig.~\ref{fig:fig2}(c)]. These effects can be further mediated by use of more advanced DD schemes~\cite{Sagi2010_Process,Almog2011}. Once the DD pulses cease there is a rapid decay followed by a revival at about a $T_{\text{osc}}$ later, in qualitative agreement with a generalization of our predictions of Eq.~\eqref{eq:coherence_Echo}.

\begin{figure}
	\centering
\begin{overpic}
[width=\linewidth]{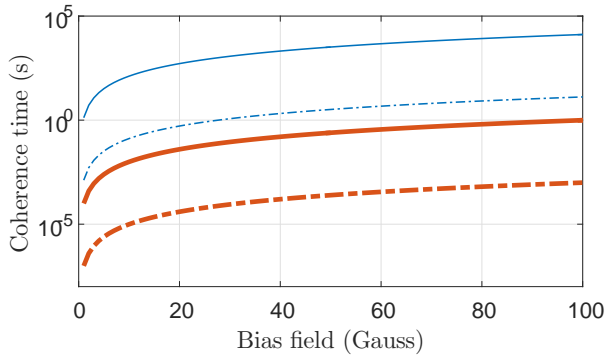}
\end{overpic}
		\caption[Coherence time in a magnetic trap]{{Predicted coherence time in a magnetic trap.} Coherence time, defined as the time at which the coherence at the $\nu^\text{th}$ revival reaches a value of 1/2, as a function of the bias field. Solid lines represent a temperature of 1~$\mu$K and dash-dotted lines a temperature of 10~$\mu$K. Red color (thick) represents an angle between the Raman beams of $180^o$, typical for atomic interferometers and blue color (thin) represents a $1^o$ angle, typical for atomic memories. For cold atomic ensembles and small angles the coherence is maintained for as long as $\sim10^4$~s. This analysis isolates the effect of anharmonicity and disregards other decoherence mechanisms.}
\label{fig:fig6}
\end{figure}

Let us now analyze a simple case of a realistic trapping potential with weaker anharmonicity. Specifically, a Ioffe-Pritchard magnetic trap, commonly used in cold atoms. In this trap, the anharmonicity is simply controlled by a magnetic field bias from the trap bottom~\cite{Esslinger1998,Steinhauer2002}. We calculate the trap ensemble anharmonicity $\mathcal{A}$ as a function of the bias magnetic field for realistic parameters and present the resulting Raman coherence time, defined as the time at which the coherence revival of Eq.~\eqref{eq:error} drops to 1/2~\footnote{For the magnetic trap we use the expansion to the $4^\text{th}$ order in $x$ of the magnetic field magnitude $|\vec{B}|=\sqrt{B_x^2+B_\text{bias}^2}$ for the $y=0$ plane with $\hat{z}$ pointing along the axis of the bias coil. Here $B_x$ is linear in $x$, $B_\text{bias}$ is the bias field and we use the \Rb wavelength $\lambda=780$~nm. For the correlation time $t_c$ we use the expression given in Eq.~\eqref{eq:error} and solve for $\epsilon_\nu=1/2$. Defining $t_c=\nu(\epsilon_\nu=1/2)T_\text{osc}$ we obtain $t_c=\left(6\pi\omega_\text{osc}\mathcal{A}^2\mathcal{N}^2\right)^{-1}$}. The obtained coherence times, isolating the effect of anharmonicity, are presented in Figure~\ref{fig:fig6}. The coherence time ranges from sub-$\mu$s for 10~$\mu$K atoms at a small bias (representing high anharmonicity) and a large angle between the beams ($180^o$, typical for atomic interferometers~\cite{Horikoshi2007,Burke2008,Segal2010,Kafle2011,Leonard2012,Fogarty2013,Mizrahi2013}) to $\sim10^4$~s for 1~$\mu$K atoms, strong bias field and a small angle ($1^o$, typical for atomic memories~\cite{Zhao2009Long,dudin2010light,dudin2013light}). Echo pulses given at trapping periods will increase these times even further~\footnote{We note that a revival of coherence for a large number of trap oscillations has been observed with a single ion in a linear Paul trap~\cite{Mizrahi2013}, however the frequency there is high, on the order of tens of MHz, leading to an overall shorter coherence time.}.

\subsection*{Summary and outlook}

In this paper we have demonstrated experimentally and analyzed theoretically and numerically the effect of a trapping potential on Raman-imprinted atomic coherence. Clear revivals of the coherence at integer multiples of the trapping period have been predicted and measured. We have shown that the amplitude of the observed revivals is highly affected by anharmonicity of the trapping potential. In addition, we predict that for traps with weak anharmonicity, quantum control methods such as echo and dynamical decoupling can increase the revival amplitude, thereby mitigating the deterioration of the coherence due to the trap anharmonicity and elastic atomic collisions. Raman atomic coherence in a trap is closely related to the field free-oscillation atom interferometry~\cite{Horikoshi2007,Burke2008,Segal2010,Kafle2011,Leonard2012,Fogarty2013} and may help to further analyze the limitations and properties of such interferometers.

Our results apply also to light storage~\cite{zhao2009millisecond,schnorrberger2009electromagnetically, dudin2010light, dudin2013light}, where the stored atomic coherence is released in the form of light emitted in a controlled direction. We expect that this type of experiment would generate similar results to the one presented in this paper, \ie a fast reduction in the power of the diffracted light, a partial revival after the trapping period and improved signals when spin echo or dynamical decoupling are used. Such stored-light experiment can be used to store images~\cite{Shuker2008} whose number of resolution pixels correspond to $\mathcal{N}$, the number of fringes used in Eqns.~\eqref{eq:coherence_harmonic} and~\eqref{eq:bad_echo}.

\appendix*

\section{Assessment of the trap anharmonicity}\label{sec:appendix}

\begin{figure}
	\centering
\begin{overpic}
[width=\linewidth]{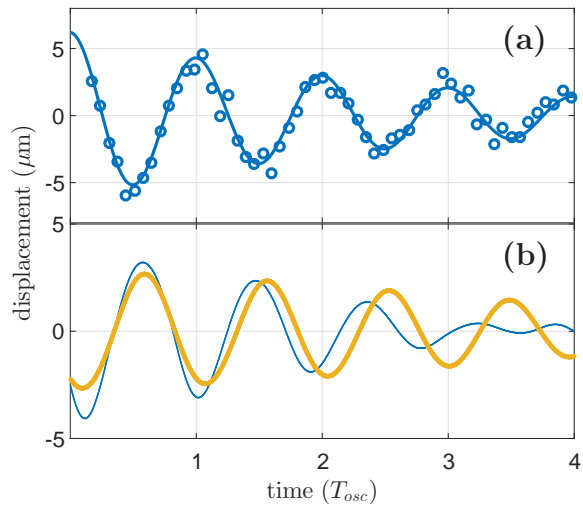}
\put(195,174){\large \textbf{(a)}}
\put(195,92){\large \textbf{(b)}}
\end{overpic}
		\caption[Calibrated anharmonicity]{{Measurement of the trap anharmonicity.} {\bf (a)} Center of mass trap oscillations, excited by giving a small kick to the trapped atoms and imaging their subsequent position after a variable oscillation time in the trap. Solid curve is the fit to a sine function with an exponentially decaying envelope, giving approximately 2.7 oscillations ($\sim9$~ms) until $1/e$ decay. This corresponds to an ensemble anharmonicity of 0.4. {\bf (b)} Simulation results of trap oscillations in a ideal cross-Gaussian beam trap (orange, thick) and a ideal cross-Gaussian beam trap superimposed with a Laguerre-Gaussian beam of calibrated intensity (blue, thin). The number of oscillations before a $1/e$ decay is 4.8 for the prefect cross-Gaussian beam trap (corresponding to an ensemble anharmonicity of 0.2) and 2.7 for the superimposed trap, in agreement with the measured data of (a).}
\label{fig:figA1}
\end{figure}

An assessment of the anharmonicity of our trap is obtained by independently measuring the decay dynamics of the center of mass trap oscillations, excited by giving a small kick to the trapped atoms. Fig.~\ref{fig:figA1}(a) depicts such an experiment, performed under the same experimental conditions described previously. It reveals a decay of the oscillation amplitude to $1/e$ after approximately 2.7 oscillations ($\sim9$~ms), Corresponding to $\mathcal{A}\approx0.4$. We use this data to extract the anharmonicity of our trap, by simulating an ideal cross-Gaussian beam trap and superimposing it with a Laguerre-Gaussian beam of calibrated intensity [Fig.~\ref{fig:figA1}(b), thin blue] to give a similar decay. For the ideal cross-Gaussian beam trap [Fig.~\ref{fig:figA1}(b), thick orange], we obtain the significantly longer decay of 4.8 oscillations.

\begin{acknowledgments}
GA and JC contributed equally to this work. 
The authors would like to thank Chen Avinadav and Arnaud Courvoisier for careful reading of the manuscript and helpful comments and suggestions. 
This work was partly supported by the Weizmann Institute Texas A\&M collaboration program.
\end{acknowledgments}

\bibliographystyle{apsrev4-1}
\bibliography{Raman_Revival}

\end{document}